\documentclass[10pt]{article}
\usepackage{amssymb}
\usepackage{amsmath}
\usepackage{latexsym}
\newcommand{\bea}{\begin{eqnarray}}
\newcommand{\eea}{\end{eqnarray}}
\newcommand{\be}{\begin{equation}}
\newcommand{\ee}{\end{equation}}



\begin{document}
\title{
\vspace{0.1in} {\bf\Large On the Classifications of Scalar Evolution Equations with Non-constant Separant}
\author {
        {\bf  Ay\c{s}e H\"{u}meyra B\.ILGE}\\
        {\it \small Faculty of Engineering and Natural Sciences, Kadir Has University, Istanbul, Turkey}\\
        {\it \small e-mail: ayse.bilge@khas.edu.tr}\\
        {\bf  Eti M\.IZRAH\.I}\\
        {\it \small Department of Mathematical Engineering, Istanbul Technical University,  Istanbul, Turkey}\\
        {\it \small e-mail: mizrahi1@itu.edu.tr}\\ {} \\
        }
}\date{\vspace{-10ex}}
        \maketitle \vspace{1mm}
\begin{center}
	{\bf Abstract}
	\vspace{3mm} \hspace{.05in}\parbox{4.5in} {{\small
			The ``separant" of the evolution equation $u_t=F$, where $F$ is some differentiable function of the derivatives of $u$ up to order $m$ is the partial derivative ${\partial F}/{\partial u_m},$ \ where $u_m={\partial^m u}/{\partial x}^m$.
			We apply the formal symmetry method proposed in [A.V.Mikhailov, A.B.Shabat and V.V.Sokolov, in V.E. Zakharov, Springer-Verlag, Berlin, (1991)]  to the  classification of   scalar evolution equations of orders $m\le 15$,  with non-trivial $\rho^{(-1)}=\left[\partial F/\partial u_m\right]^{-1/m}$ and   \ $\rho^{(1)}$.
			We obtain the   ``top  level" parts of these equations and   their ``top dependencies"   with respect to the ``level grading" defined in [E.Mizrahi, A.H.Bilge, J. Phys. A: Math. Theor. 46 (2013),  385202].
			We show that if $\rho^{(-1)} $ depends on $u,u_1,\dots,u_b,$ \ where $b$ is the base level, then, these equations are level homogeneous polynomials   in $u_{b+i},\dots ,u_m$, $i\ge 1$   and the coefficient functions are determined  up to their dependencies on $u,u_1,\dots,u_{b-1}$.
			We prove that if $\rho^{(3)}$ is non-trivial, then
			$\rho^{(-1)}=(\alpha u_b^2+\beta u_b+\gamma)^{1/2}$, with $b\le 3$ while if
			$\rho^{(3)}$ is trivial, then
			$\rho^{(-1)}=(\lambda u_b+\mu)^{1/3}$, where $b\le 5$  and  $\alpha$, $\beta$, $\gamma$,   $\lambda$ and  $\mu$  are functions of $u,\dots,u_{b-1}$.
		  We  show that these equations form  commuting flows and we construct   their recursion operators that are respectively of orders $2$ and $6$ for non-trivial and trivial $\rho^{(3)}$ respectively.
			Omitting lower order dependencies,  we show that
 equations with non-trivial $\rho^{(3)}$ and $b=3$ are symmetries of the ``essentially non-linear third order equation".  For trivial $\rho^{(3)}$, the equations with $b=5$  are symmetries of a non-quasilinear fifth order equation obtained in [A.H.Bilge, Computers and Mathematics with Applications,
			{\bf 49}, (2005), 1837-1848] while for $b=3,4$ they are symmetries of quasilinear fifth order equations and
 we outline the transformations to polynomial equations where $u$ has zero  scaling weight, suggesting that the hierarchies that we obtain could be
			 transformable to known equations  possibly by introducing non-locality.
		}}
	\end{center}
\noindent
{\it \footnotesize 2010 Mathematics Subject Classification}: {\scriptsize 35Q53, 37K10}.\\
{\it \footnotesize Key words}: {\scriptsize classification,
differential polynomials, evolution equations, hierarchies, }
\maketitle

\section*{1. Introduction}

The term ``integrable equations" refers to those equations that are either linearizable or  solvable by the inverse spectral transformation \cite{gardner}.  The prototype of integrable equations is the Korteweg-deVries (KdV) equation  characterized, among others, by an infinite sequence of symmetries, an infinite sequence of  conserved densities and  a recursion operator. The search for integrable equations led to the discovery of a number of equations that are transformable to the KdV equation by the so-called Miura  transformations and to two new hierarchies called the Sawada-Kotera \cite{sawada}  and Kaup-Kupershmidt \cite{kaup}  equations.
KdV hierarchy starts at order $m=3$ and have symmetries at all odd orders, while the  Sawada-Kotera and Kaup equations belong to hierarchies starting at order $m=5$ and  have symmetries  at odd orders that are not multiple of $3$.

The search for new integrable hierarchies was marked by  negative results of
 Wang and Sanders \cite{SW98},  proving that polynomial equations of order $m\ge 7$ are symmetries of lower order equations and extending this result  to certain types of non-polynomial equations \cite{SW2000}. We aimed to obtain a similar uniqueness result for general, non-polynomial integrable equations and we applied
the  method of formal symmetries, based on the existence of ``canonical densities" \cite{MSS1991},
 to the classification of
 integrable evolution equations in $1+1$ dimensions, $u_t=F[u]$   \cite{B2005,MB2009,OB2012,MB2013}.

 In \cite{B2005}, we obtained the canonical densities  $\rho^{(i)}$,  $i=1,2,3$  for evolution equations of order $m\ge 7 $ and we have shown that evolution equations admitting a conserved density of order $n>m$, \  are quasi linear.  Then, in \cite{MB2009} we showed that evolution equations with non-trivial  $\rho^{(i)}$,  $i=1,2,3$  are
 polynomial in $u_{m-1}$ and $u_{m-2}$  and possess  a certain scaling property that we called ``level grading" \cite{MB2013}.
 For $m=5$, we have shown that  there is a candidate for non-quasilinear integrable equation \cite{B2005}, we    obtained canonical densities $\rho^{(i)}$,  $i=1,\dots ,5$ and obtained a preliminary classification of quasilinear  $5$th order equation with non-trivial $\rho^{(i)}$,  $i=1,2,3$ \cite{OB2012}.

In subsequent work on the classification problem, we noticed that the triviality or non-triviality of the canonical density $\rho^{(3)}$  is the key element that determines the form of  integrable equations. In analogy with the fact that
 the KdV equation has conserved densities of all orders while the hierarchies of Sawada-Kotera and Kaup equations have missing conserved densities, we called those equations that admit an unbroken sequence of conserved densities as ``KdV-type" and those for which $\rho^{(3)}$ is trivial as ``Sawada-Kotera-Kaup-type".
In preliminary work,  we considered also evolution equations with non-constant but trivial $\rho^{(-1)}$ and/or trivial $\rho^{(1)}$, but these cases turned out to be difficult to deal with and they  were disregarded in subsequent analysis.

 In the present work, we consider scalar evolution equations in $1$ space dimension, $u_t=F(u,u_1,\dots, u_m)$, $m\le 15$ and we  obtain their classifications up to  ``Top Levels" and ``Top Dependencies" (to be defined in Section 2), using the ``Formal Symmetry" method \cite{MSS1991}, assuming that the canonical densities $\rho^{(-1)}$ and $\rho^{(1)}$ are non-trivial.
 We recall that
 $$\rho^{(-1)}=\left[\frac{\partial F}{\partial u_m}\right]^{-1/m}$$
 and use  the notation
 $$A=\frac{\partial F}{\partial u_m}=a^m=\left[\rho^{(-1)}\right]^{-m}.$$
 In all cases, for reasons discussed in Remark 2, we assume that
 $$\partial \rho^{(-1)}/\partial u_3\ne 0.$$

 The results can be summarized as below.
 If $\rho^{(3)}$ is non-trivial, then integrable equations are polynomial in $u_k$, $k \ge 4,$ and $a=1/\rho^{(-1)}$, where $\rho^{(-1)}$ has the form
 $$\rho^{(-1)}=(\alpha u_3^2+\beta u_3+\gamma)^{1/2}.\eqno(1.1)$$
On the other hand, if  $\rho^{(3)}$ is trivial, then integrable equations are non-trivial only at odd  orders  that are not multiples of $3$; they are polynomial in $u_k$, $k\ge b$, where  $b=3,4,5$ and    $a=1/\rho^{(-1)}$, and   $\rho^{(-1)}$ has the form
$$\rho^{(-1)}=(\lambda u_b+\mu)^{1/3},\quad b=3,4,5.\eqno(1.2)$$
In all cases, we prove that  the equations that we obtain form commuting flows and  we construct  their recursion operators that are of orders $2$ and $6$ respectively
for the KdV and Sawada-Kotera-Kaup types.
Finally we uses series of potentiations to convert the non-quasilinear equations to quasilinear ones, omitting lower order dependencies.

In Section 2, we recall basic definitions and  ``Level Grading". In Section 3, we describe our solution procedure and present the results in Section 4. The construction of the recursion operator and  explicit forms of the flows are given in Section 5.    Discussion of the results and a brief outline of the transformations to known equations are given  in Section 6.

\section*{2. Preliminaries}

\vskip 0.2cm
\noindent
{\bf Basic Definitions }
\vskip 0.2cm

We consider evolution equations in $1+1$ dimension of order $m$. The unknown function is $u(x,t)$ and the $n$ th derivative of $u$ with respect to $x$ is denoted as $u_n$.  The evolution equation is of the form   $u_t=F[u]$, where $F$ is some function of $u$, $u_1$, $\dots $, $u_m$.
A ``symmetry" $\sigma$, of the evolution equation  is a solution of the linearized equation $\sigma_t=F_*\sigma$, where $F_*=\sum_i \partial F/\partial u_i D^i$ is the Frechet derivative of $F$, i.e,
$$\sigma_t=F_*\sigma= \sum_i \frac{\partial F}{\partial u_i} D^i(\sigma).\eqno(2.1)$$
A ``conserved covariant" or a ``co-symmetry" $\gamma$, is a solution of $\gamma_t=-F_*^\dagger\gamma$, where $F_*^\dagger$ is the adjoint of $F_*$, i.e,
$$\gamma_t=-F_*^\dagger\gamma=\sum_i (-1)^{i+1}  D^i\left( \frac{\partial F}{\partial u_i}   \gamma  \right).\eqno(2.2)$$
 A ``conserved density" $\rho$  satisfies $\rho_t=D\eta$, for some $\eta$, that is
$$\rho_t=\sum_i   \frac{\partial \rho}{\partial u_i} D^i F=D\eta.\eqno(2.3)$$
Using identities involving variational derivatives, it can be shown that \cite{MSS1991}, the variational derivative $\frac{\delta \ }{\delta u}$  of a conserved density is a  conserved covariant
 $$\gamma=\frac{\delta \rho }{\delta u}= \sum_i (-1)^i \frac{\partial \rho}{\partial u_i}.\eqno(2.4)$$

\vskip 0.2cm
\noindent {\bf Remark 1.}
If a conserved co-variant $\gamma$ is the variational derivative of a conserved density, then  $\gamma F $ is a total derivative,
 because if $\rho$ is a conserved quantity, then
$$\rho_t=\sum \frac{\partial \rho}{\partial u_i}D^i F= \sum (-1)^i D^i \left(\frac{\partial \rho}{\partial u_i}\right) F=\gamma F=D\eta.\eqno(2.5)$$

\vskip 0.2cm
\noindent
{\bf The Recursion Operator}

 A  ``recursion operator" is defined as an integro-differential operator that  sends symmetries to symmetries, i.e,
 $R\sigma$ should be a symmetry whenever  $\sigma $ is \cite{O77}. In particular,
 	  $R\sigma$ has to be local function.
It can be easily seen that if $\sigma$ and $R\sigma$ are both symmetries, then,
 $$(R_t+[R,F_*])\sigma =0.\eqno(2.6)$$
 A ``formal symmetry" is a Laurent series that satisfies $R_t+[R,F_*]=0$  up to a certain order.
The solvability of $R_t+[R,F_*]=0$  in the class of local functions is   equivalent to the locality of the time evolution of certain quantities called the ``canonical densities". The canonical densities are computable in terms of the partial derivatives of $F$ and their conservation is proposed as an integrability test in \cite{MSS1991}.

\vskip 0.2cm
\noindent
In this work we will assume that the canonical density $\rho^{(-1)}=\left(\partial F/\partial u_m\right)^{-1/m}$ and $\rho^{(1)}$ are   nontrivial.
The triviality or non-triviality of $\rho^{(3)}$  distinguishes between two classes of integrable equations that we called  ``KdV-type" or ``Sawada-Kotera-Kaup-type.

The derivation of the classification results depends heavily on the properties of ``level grading".

\vskip 0.2cm
\noindent
{\bf The ``Level Grading":}

 Let $K$ be the ring of differentiable functions of $u,u_1,\dots u_k$.  The module generated by $u_{k+1}, u_{k+2},\dots$ has a graded algebra structure. We called this grading as the ``level" of polynomials in $u_{k+j}$ above the base level $k$.

 As it can be easily checked, differentiation increases the level by $1$. For example,
 if $\varphi=\varphi(x,t,u,\dots,u_k)$, then
$$D\varphi= \underbrace{\frac{\partial\varphi}{\partial u_k}u_{k+1}}_{\rm level \ 1}+ \alpha,\quad
D^2\varphi=\underbrace{\frac{\partial\varphi}{\partial u_k} u_{k+2}
   +\frac{\partial^2\varphi}{\partial u_k^2} u_{k+1}^2}_{\rm level \ 2}
+\underbrace{\beta u_{k+1}}_{\rm level \ 1}+\gamma,$$
where $\alpha$, $\beta$, $\gamma$ are certain expressions that  depend on at most $u_k$.
 The crucial property that makes the level grading a useful tool is its invariance under  integration by parts.
  Let $k<p_1<p_2<\dots<p_l<s-1$. Integration by parts applied to monomials linear in the highest derivative gives either of the forms below.
  \begin{eqnarray*}
  \varphi u_{p_1}^{a_1}\dots u_{p_l}^{a_l}u_s&\cong& -D\left(\varphi
  u_{p_1}^{a_1}\dots u_{p_l}^{a_l}\right) u_{s-1}, \nonumber\cr
  \varphi u_{p_1}^{a_1}\dots u_{p_l}^{a_l}u_{s-1}^pu_s&\cong&
  -\textstyle\frac{1}{p+1} D\left(\varphi u_{p_1}^{a_1}\dots
  u_{p_l}^{a_l}\right)  u_{s-1}^{p+1}.\nonumber\cr
  \end{eqnarray*}
  It can be seen that the level above $k$ is preserved in both cases.
  Integrations by parts are repeated until one encounters a
  non-integrable monomial such as
  $$u_{p_1}^{a_1}\dots u_{p_l}^{a_l}u_{s}^p,\quad p>1,$$
  which has the same level as the original expression.

If $P$ is a polynomial in $u_k$, $k>b$, with coefficients depending on $u_i$, $i\le b$, then the monomials in $P$ can be arranged according to their levels above the base level $b$.  The parts that has the highest level is called the  ``Top Level" part of $P$ and the dependency of the coefficients of the top level on $u_b$ is called their ``Top Dependency"\cite{MB2013}.

\section* {3. Solution Procedure}

We recall that, the classification problem for  polynomial integrable equations is solved by Wang and Sanders \cite{SW98},  by proving that integrable polynomial evolution equations of order $m\ge 7$ are symmetries of a lower order equation.  With the aim of obtaining a similar result, we have undertaken a program for classifying non-polynomial evolution equations by the formal symmetry method.  The first result in this direction was the quasi-linearity, obtained in \cite{B2005}. We have proved the following.

\vskip 0.2cm
\noindent
{\bf Proposition 1.}  {\it Assume that the evolution equation  $u_t=F(u,u_1,\dots, u_m)$, with $m\ge 7$,  admits a conserved density of order $n=m+1$. Then $F$ is linear in $u_m$.}
\vskip 0.2cm
This result is not valid for  $m=5$, as there is a non-quasilinear integrable evolution equation of order $5$, as it will be discussed in the next section.

Although we have obtained polynomiality in top $3$ derivatives in \cite{MB2009},
we start here with the quasilinear form and indicate the steps towards the classification of lower order evolution equations.

Since we deal with quasilinear equations, the separant is the coefficient of $u_m$, we denote it either $A$ and parametrize it as $A=a^m$.
 We start with the quasilinear case,
$$u_t=A u_m +B,\eqno(3.1)$$
 where $A$ and $B$ depend on the derivatives of $u$ up to $u_{m-1}$. As we work with top level  terms, we  let $F=Au_m$ and we assume only top dependency i.e, we let $A=A(u_{m-1})$.
Then  we use the conservation laws for the canonical densities $\rho^{(-1)}$ and $\rho^{(1)}$ to get $\partial A/\partial u_{m-1}=0$.  We finally show that $\partial^3B/\partial u_{m-1}^3=0$, hence evolution equation has the form
$$u_t=Au_m+Bu_{m-1}^2 +Cu_{m-1}+E,\eqno(3.2)$$
where the coefficients are functions of the derivatives of $u$ up to order $m-2$.
As before,  the conservation laws for $\rho^{(-1)}$ and $\rho^{(1)}$ are used to solve the coefficient functions $B$ and $C$ in terms of the derivatives of $a$. For $m\ge 9$ we obtain $\partial a/\partial u_{m-2}=0$. For  $m=7$, if $\rho^{(3)}$ is trivial, then $a$ has a non-trivial dependency on $u_5$, but if $\rho^{(3)}$ is non-trivial, we get $a_5=0$.
In the case where $a_{m-2}=0$, the top level part is linear and we show that $F$ is a sum of level homogeneous terms of level $3$ above the base level $m-3$, as given below.
$$u_t=Au_m +Bu_{m-1}u_{m-2}+Cu_{m-1}^3+Eu_{m-2}+Gu_{m-1}^2+Hu_{m-1}+ K.\eqno(3.3)$$
In this case also, for $m\ge 9$, we can show that the separant is independent of $u_{m-3}$, the top level part is linear and equation is of level $4$ above base $m-4$. For $m=7$, the base level is $m-3=4$; if $\rho^{(3)}$ is trivial, there are equations with non-trivial separant but if $\rho^{(3)}$ is non-trivial, then the evolution equation is of level $4$ above the base level $3$.

 $F$ is now  a sum of level homogeneous terms of level less than or equal to $4$ above the base level  $m-4$. At this stage, it is more convenient to switch to the notation,
$$u_m=u_{b+4},\quad u_{m-1}=u_{b+3},\quad  u_{m-2}=u_{b+2},\quad u_{m-3}=u_{b+1},\quad u_{m-4}=u_{b},\eqno(3.4)$$
and write $u_t$ as
\begin{align*}
u_t= Au_{b+4} &+Bu_{b+3}u_{b+1}+Cu_{b+2}^2+Eu_{b+2}u_{b+1}^2 +Gu_{b+1}^4\\
&+ Hu_{b+3}+Iu_{b+2}u_{b+1} +Ju_{b+1}^3
+Ku_{b+2}+Lu_{b+1}^2 \\
&+Mu_{b+1}+N.\tag{3.5}
\end{align*}
For $m=7$, the base level is $3$ and $\partial A/\partial u_3$ is non-zero regardless of the triviality of $\rho^{(3)}$.  The functional form of $A=(\rho^{(-1)})^{-m}$  depends on the non-triviality or triviality of $\rho^{(3)}$, as given respectively by equations (1.1) and (1.2). For $m\ge 9$, the base level is $5$; the conserved density conditions imply that $a_5=0$  regardless of the triviality of  $\rho^{(3)}$.  For $m\ge 11$,  $\rho^{(-1)}$ and $\rho^{(1)}$ only imply that $a_b=0$, $b=m-4$.  It follows that $u_t$ is a level homogeneous polynomial of level $b$ above the base level $b=m-5$.

For each order $m=2k+1$, $ 9\le m\le 15$,  and  base level $b\le m-5$, we solve the conserved density conditions in a similar manner, the form of the evolution equation depending on the difference $m-b$ only.

%

The form of polynomials of level $j=m-b$ are obtained using  partitions of the integer $j$. For low values of $j=m-b$, these forms  can be found by inspection, but for higher values of $j$, the number of partitions grow quickly. We used an algorithm implemented by
J. Errico \cite{Errico}.  The explicit form of the partitions of integer matrix is given in the Appendix A.  The passage from the matrix of the partition of the integer $j$  to the polynomial of level $j$ above the base level $b$ is achieved by a REDUCE program as described in Appendix B, where we also present the results for $j\le 10$.

In the case where $\rho^{(3)}$ is non-trivial, its explicit form is not needed.
The  generic form of the conserved densities  depends on the base level $b$ only. They are of the form
$$
\rho^{(-1)}=a^{-1},\quad
\rho^{( 1)}=P^{(1)}u_{b+1}^2,\quad
\rho^{( 3)}=P^{(3)}u_{b+2}^2+Q^{(3)}u_{b+1}^4,\eqno{(3.6)}
$$
where $P^{(i)}$, $Q^{(i)}$ depend on $u_b$.
The computation of the integrability conditions depends on the order $m=2k+1$, as follows.  As we use only the top dependency, i.e, the dependency on $u_b$, the time derivatives are
\begin{align*}
	\rho^{(-1)}_t=&(a^{-1})_bD^bF,\\
	\rho^{( 1)}_t=&2P^{(1)}u_{b+1}D^{b+1}F+P^{(1)}_bu_{b+1}^2D^bF,\\
	\rho^{( 3)}_t=& 2P^{(3)}u_{b+2}D^{b+2}F
	                +4Q^{(3)}u_{b+1}^3D^{b+1}F\\
	&\quad\quad+(P^{(3)}_bu_{b+2}^2+Q^{(3)}_bu_{b+1}^4)D^{b}F .\tag{3.7}
\end{align*}
These time derivatives are evaluated  for each order  and each base level using REDUCE programs. A sample  program is given  in Appendix C.  For higher orders, we had to use also  conserved density conditions for a generic expression of $\rho^{(5)}$, as level homogeneous polynomial of level $8$ above the base level  $b$.

For base levels $b>5$,  the conserved density conditions imply that $\partial A/\partial u_{b}=0$ and we show furthermore that the flow is polynomial in $u_b$.  For $b\le 5$, the conserved density conditions for  $\rho^{(-1)}$ and $\rho^{(1)}$
should be supplemented by the information on whether  $\rho^{(3)}$ is trivial or not.

For the case with non-trivial $\rho^{( 3)}$ we obtain the dependencies of the flows of order $m\le 15$ on $u_k$, $k\ge 3$, as presented in the next section together with the recursion operator.  These equations are polynomial in $u_k$, $k\ge 4$ and their $u_3$ dependency is via $ a$ and its derivative with respect to $u_3$.  The dependencies on $u_k$, $k=2,1,0$ are omitted. This hierarchy is characterized by the form of $a$ given by Eq.(1.1),
$$a=(\alpha u_3^2+\beta u_3+\gamma)^{-1/2}.\eqno(3.8)$$

For the case where $\rho^{(3)}$ is trivial, we use its explicit expression and set the quantities $P^{(3)}$ and $Q^{(3)}$ to zero and add these to the constraints imposed by the conserved density conditions for $\rho^{(-1)}$ and $\rho^{(1)}$
 For $b=5$ and $b=4$, the equations with non-trivial separant occur only for $\rho^{(3)}$ trivial while for $b=3$, we have a non-trivial separant regardless of the triviality of $\rho^{(3)}$. The flows are polynomial on $u_{b+j}$ and the dependency on $u_b$ is via
$$a=(\lambda u_b+\mu)^{-1/3},\quad b=3,4,5\eqno(3.9)$$
and its derivatives. The first few flows are presented in the next section together with their recursion operators.

\vskip 0.2cm
\noindent
{\bf Remark 2.} If the base level is $1$, that is if the separant or equivalently $\rho^{(-1)}$ depends on $u$ and $u_1$ only, then one can use transformations of type (1.4.13,14) given in \cite{MSS1991} to transform the separant to $1$.  When  $b= 2$ and $\rho^{(3)}$ is non-trivial we get $a$ in the form of (3.8) with $b=2$, but when $\rho^{(3)}$ is trivial, we obtain a  third order ordinary differential equation for $a$, which admits (3.9) as a special solution.
Since the candidates of integrable equations for $b=2$ form possible a larger class, we omit this case in the present work and we assume that $b\ge 3$.
\vskip 0.2cm

\section*{4. Results}

\subsection*{Order $3$}
    Evolution equations of order $3$ are classified in \cite{MSS1991}. These fall in $3$ classes (3.3.7-9).
    \begin{align*}
    u_t=&A_1u_3+A_2,\tag{4.1a}\\
    u_t=&(A_1u_3+A_2)^{-2}+A_3,\tag{4.1b}\\
    u_t=&(2A_1u_3+A_2)(A_1u_3^2+A_2u_3+A_3)^{-1/2}+A_4.\tag{4.1c}
    \end{align*}
We note that here, unlike our convention, subscript refer to indices, not to differentiations.  The first equation is quasilinear.  The second one is characterized by the triviality of $\rho^{(-1)}$, therefore it is excluded from our discussion. The third equation is known as the essentially non-linear third order equation studied further in \cite{HSS95}.

When we start with the essentially non-linear equation in the form above, we see that
$\rho^{(-1)}$ is
$$(2A_1A_3-{\textstyle\frac{1}{2}}A_2^2)^{-1/3}  (A_1u_3^2+A_2u_3+A_3)^{1/2}.\eqno(4.2)$$
This is not exactly the form that we want.  In order to obtain $\rho^{(-1)}$ in the same form as the higher order equations, i.e, in the form (1.1),  we should start with the evolution equation
$$u_t=(-{\textstyle\frac{1}{2}}\beta^2+2\alpha\gamma)^{-1}
(\alpha u_3^2+\beta u_3+\gamma)^{-1/2}(2 \alpha u_3 +\beta)+\delta.\eqno(4.3)$$
We now prove that this choice is possible. In \cite{B2005}, Proposition 4.4, we have shown that if $\rho^{(-1)}=F_m^{-1/m}$ is conserved, then $\rho^{(-1)}_{m,m}\ne 0$, then
	$$F_m=\left[ c^{(1)}F^2+     c^{(2)}  F +    c^{(3)}    \right]^{m/(m-1)}.\eqno(4.4)$$
It can be shown that the  forms (4.1c) and (4.3)   are both consistent with this equation, thus we can start with the form (4.3) that we write as
$$u_t={\textstyle\frac{4}{P}} a^{-2} a_3, \quad \quad  P=\beta^2-4\alpha\gamma,\eqno(4.5)$$
at the top level and
$$a=(\alpha u_3^2+\beta u_3+\gamma)^{-1/2}\eqno(4.6)$$
Thus, the essentially nonlinear third order equation is characterized by
 $$\rho^{(-1)}=(\alpha u_b^2+\beta u_b+\gamma)^{1/2},\eqno(4.7)$$
with $b=3$.

 \subsection*{Order $5$}

If $\rho^{(3)}$ is non-trivial, than the top level part of integrable equations of order $5$ is of the form
$$u_t=a^5 u_5 +5/2 a^4 a_3 u_4^2.\eqno(4.8)$$
This equation is characterized by $\rho^{(-1)}$ of the form (4.7) above.  We have in fact checked that it is a symmetry of the essentially nonlinear third order equation (4.5).
In \cite{OB2012}, we have obtained lower order terms and we presented  a special solution.

For the case where $\rho^{(3)}$ is trivial we have $3$ classes of solutions with base levels $b=5,4,3$.
In \cite{B2005}, we have shown that for $m=5$ there is a candidate of integrable equation of the form
$$u_t=-\frac{3}{2\lambda}(\lambda u_5+\mu)^{-2/3}+\nu\eqno(4.9)$$
where $\lambda$, $\mu$ and $\nu$ are independent of $u_5$.  This form can also be obtained from the triviality of $\rho^{(3)}$, whose explicit expression is given in \cite{OB2012}. This expression involves $\int \rho^{(-1)}_t$, hence we first compute it up to some unknown function depending on at most $u_5$. Then we substitute this in the expression of $\rho^{(3)}$. The coefficient of $u_7^2$ gives
$\partial^2 a/\partial u_5^2=  4 a_5^2 a^{-1}$.
This equation can  be integrated twice to give
$$a=(\lambda u_5+\mu)^{-1/3}.\eqno(4.10)$$
 Expressing $\lambda$ in terms of $a_5$ we obtain the alternative form
$$u_t=\frac{1}{2}\frac{a^6}{a_5}.\eqno(4.11)$$
Continuing with trivial $\rho^{(3)}$ and assuming  $\frac {\partial a}{\partial u_5}=0$, we obtain  $F$ simply as
$$u_t=a^5 u_5,\eqno(4.12)$$
with
$$a=(\lambda u_4+\mu)^{-1/3}.\eqno(4.13)$$
Finally, again with trivial $\rho^{(3)}$ and  $\frac {\partial a}{\partial u_4}=0$, we obtain
$$u_t=a^5 u_5+5 a^4a_3 u_4^2,\eqno(4.14)$$
with
$$a=(\lambda u_3+\mu)^{-1/3}.\eqno(4.15)$$
 By Remark 2, we omit the cases for $b\le 2$.

\subsection*{The Hierarchy Structure }

We obtained the explicit forms of integrable equations for $m\le 15$ and $b=3,4,5$ using REDUCE interactively as outlined in the Appendices. Furthermore, we also compute flows of order $m=17$ as a symmetry of $5$th order equations.
We have explicitly checked that all evolution equations with non-trivial     $\rho^{(3)}$ are symmetries of the essentially nonlinear third order equation (4.5) and they form a commuting flow. Similarly we have explicitly checked that equations with trivial $\rho^{(3)}$ over base levels $b=3,4,5$ are symmetries of $5$th order equations (4.14), (4.12) and (4.11) respectively and they form commuting flows.

\section*{ 5. Construction of the recursion operator}

In \cite{B1993}, we have shown that if the recursion operator has the form
$$R=R^{(n)}D^n+R^{(n-1)}D^{n-1}+\dots +R^{(1)} D+ R^{(0)}+\sum_{i=1}^N \sigma_iD^{-1}\gamma_i,\eqno(5.1)$$
then  $\sigma_i$ has to be  a symmetry and $\gamma_i$ has to be a conserved covariant. But in general, there is no guarantee that the recursion operator will have a {\it finite}  expansion of this type.

By Remark 1, if the conserved covariants $\gamma_i$'s are chosen as variational derivatives of conserved densities, then it will follow that $R(F)$ will be a local function.
The form of the (least) order of the recursion operator can be guessed by considering the orders of the symmetries in the hierarchy
and by level grading  arguments.
Based on the form of the recursion operators for the KdV hierarchy  and Sawada-Kotera and Kaup hierarchies \cite{B1993}, we start with recursion operators of the orders $2$ and $6$ respectively.
For the KdV type equations, the recursion operator is proposed as
$$R^{(2)}D^2+R^{(1)}D+R^{(0)}+\sigma D^{-1}\gamma,\eqno(5.2)$$
where $\sigma$ is  proportional to the third order essentially non-linear equation and $\gamma$ is  the variational derivative of $1/a$.
For the Sawada-Kotera-Kaup type equations, we start with
\begin{align*}
R=a^6 D^6&+ R^{(5)}D^5 + R^{(4)}D^4 + R^{(3)}D^3 + R^{(2)}D^2 + R^{(1)}D + R^{(0)}\\
&+\sigma^{(1)} D^{-1}\gamma^{(1)}+\sigma^{(2)} D^{-1}\gamma^{(2)},\tag{5.3}\end{align*}
where $\sigma^{(i)}$, $i=1,2$ are proportional $7$th and $5$th order flows, $\gamma^{(i)}$, $i=1,2$ are the variational derivatives of $\rho^{(-1)}$ and $\rho^{(1)}$ respectively.

We  started with the form of the recursion operators as above, where the $R^{(i)}$'s were chosen as level homogeneous polynomials so that the operators $R$  have levels $2$ and $6$ respectively,  we determined the coefficient functions from the requirement that $R$ acting on a symmetry produces the next order flow.  In this procedure, for the Sawada-Kotera-Kaup type equations we needed the  expression of the flow of order $m=17$, which was obtained as a symmetry of lower order flows.
We present the results below.

\subsection*{\bf Non-trivial $\rho^{(3)}$:}
 $$R=a^2 D^2+\left[(-a_3a)u_4\right]D+\left[(a_3 a)u_5+(3 a_3^2)u_4^2 \right]+\sigma D^{-1}\eta,\eqno(5.4)$$
 where
 $$\sigma=\frac{4}{P}a_3 a^{-2},\quad
 \eta=-D^3\left[\frac{\partial a^{-1}}{\partial u_3}\right],\quad P=\beta^2-4\alpha\gamma.\eqno(5.5)$$

The first $4$ flows for non-trivial $\rho^{(3)}$ and $b=3$  are given by
\begin{eqnarray*}
u_{t,3}&=&-\frac{4}{P} a^{-2} a_3,\\
u_{t,5}&=&  a^5 u_5 + 5/2 a^4  a_3 u_4^2,\\
u_{t,7}&=&  a^7 u_7
+14		a^6 a_3	u_6 u_4
+21/2	a^6 a_3 u_5^2
+		a^5(98 a_3^2 +35/8	P a^6) u_5	u_4^2\\&&
+       a^4 a_3 (189/2 a_3^2+399/32 P  a^6 ) u_4^4\\
u_{t,9}&=&
a^9 u_9
+27		a^8	a_3		u_8 u_4
+57		a^8	a_3		u_7 u_5
+69/2	a^8	a_3		u_6^2			\\&&
+		a^7	(360	a_3^2 +105/8	Pa^6)		u_7 u_4^2
+		a^7	(1230	a_3^2 +189/4	Pa^6)		u_6 u_5 u_4\\&&
+		a^7	(290	a_3^2 +91/8		Pa^6)		u_5^3
+330	a^6 a_3(9	a_3^2 +			Pa^6)		u_6 u_4^3	\\&&
+		a^6 a_3(6105a_3^2 +11187/16 Pa^6)		u_5^2 u_4^2 \\&&
+		a^5 (16335	a_3^4+29469/8 P a^6 a_3^2  +6699/128  P^2 a^{12}) u_5 u_4^4\\&&
+		a^4 a_3(19305/2 a_3^4+57915/16 P a^6 a_3^2+39325/256  P^2 a^{12}) u_4^6.
\end{eqnarray*}

\subsection* {Trivial $\rho^{(3)}$}

The recursion operator is chosen in the form (5.3), and the coefficients are solved for each case.  If we write $R$ by factoring $a^6$, the recursion operators for $b=3,4,5$ have the same functional form, up to constants.
\begin{align*}
R=&a^6 \left( D^6+ \tilde{R}^{(5)}D^5 + \tilde{R}^{(4)}D^4 + \tilde{R}^{(3)}D^3 + \tilde{R}^{(2)}D^2 + \tilde{R}^{(1)}D + \tilde{R}^{(0)}\right)   \\
&+\sigma^{(1)} D^{-1}\gamma^{(1)}+\sigma^{(2)} D^{-1}\gamma^{(2)}. \tag{5.6}
\end{align*}
where
\begin{eqnarray*}
	\tilde{R}^{(5)}&=&   k^{(5,1)} \  q \ u_{b+1},\\
	\tilde{R}^{(4)}&=&   k^{(4,1)} \  q \ u_{b+2}   + k^{(4,2)} q^2 \ u_{b+1},\\
	\tilde{R}^{(3)}&=&   k^{(3,1)} \  q \ u_{b+3}
	+k^{(3,2)} \  q^2 \ u_{b+2} u_{b+1}
	+k^{(3,3)} \  q^3 \ u_{b+1}^3,\\
	\tilde{R}^{(2)}&=&   k^{(2,1)} \  q    \ u_{b+4}
	+k^{(2,2)} \  q^2 \  u_{b+3} u_{b+1}
	+k^{(2,3)} \  q^2 \  u_{b+2}^2     \\&&
	+k^{(2,4)} \  q^3 \  u_{b+2} u_{b+1}^2
	+k^{(2,5)} \  q^4 \ u_{b+1}^4,\\
	\tilde{R}^{(1)}&=&
	k^{(1,1)}   \  q   \	u_{b+5}
	+ k^{(1,2)} \  q^2 \   u_{b+4}u_{b+1}
	+ k^{(1,3)} \  q^2 \	u_{b+3}u_{b+2} \\&&
	+ k^{(1,4)} \  q^3 \   u_{b+3}u_{b+1}^2
	+ k^{(1,5)} \  q^3 \	u_{b+2}^2u_{b+1} \\&&
	+ k^{(1,6)} \  q^4 \	u_{b+2}u_{b+1}^3
	+ k^{(1,7)} \  q^5 \	u_{b+1}^5,\\
	\tilde{R}^{(0)}&=&
	k^{(0,1)}  \  q   \  u_{b+6}
	+ k^{(0,2)}  \  q^2 \  u_{b+5}u_{b+1}
	+ k^{(0,3)}  \  q^2 \  u_{b+4}u_{b+2} 		\\&&
	+ k^{(0,4)}  \  q^2 \  u_{b+3}^2
	+ k^{(0,5)}  \  q^3 \  u_{b+4}u_{b+1}^2
	+ k^{(0,6)}  \  q^3 \  u_{b+3}u_{b+2}u_{b+1} \\&&
	+ k^{(0,7)}  \  q^3 \  u_{b+2}^3
	+ k^{(0,8)}  \  q^4 \  u_{b+3}u_{b+1}^3
	+ k^{(0,9)}  \  q^4 \  u_{b+2}^2 u_{b+1}^2	\\&&
	+ k^{(0,10)} \  q^5 \  u_{b+2}u_{b+1}^4
	+ k^{(0,11)} \  q^6 \  u_{b+1}^6,
\end{eqnarray*}
where the $k^{(i,j)} $'s are constants and $q=a_b/a$.

We present the explicit forms for each base level together with the first $4$ flows.

  \subsection* {Trivial $\rho^{(3)}$, $b=5$}

  \begin{eqnarray*}
  R^{(5)}&=&3  a^5  a_5  u_6,\\
  R^{(4)}&=&2 u_7 a_5 a^5 + 7 u_6^2 a_5^2 a^4,\\
  R^{(3)}&=& -u_8 a_5 a^5 - 16 u_7 u_6 a_5^2 a^4 - 42 u_6^3 a_5^3 a^3,\\
  R^{(2)}&=&  u_9 a_5 a^5 + 21 u_8 u_6 a_5^2 a^4 + 16 u_7^2 a_5^2 a^4 + 262   u_7 u_6^2 a_5^3 a^3 + 490 u_6^4 a_5^4 a^2,\\
  R^{(1)}&=&
  - u_{10}a_5a^5 - 28u_9u_6a_5^2a^4 - 51u_8u_7a_5^2a^4 -
  462u_8u_6^2a_5^3a^3 - 660u_7^2u_6a_5^3a^3 \\
         &&- 5190u_7u_6^3   a_5^4a^2 - 7560u_6^5a_5^5a,\\
 R^{(0)}&=&
  u_{11}a_5a^5 + 35u_{10}u_6a_5^2a^4 + 79u_9u_7a_5^2a^4 + 742
  u_9u_6^2a_5^3a^3 + 49u_8^2a_5^2a^4 \\
      &&+ 2700u_8u_7u_6a_5
  ^3a^3 + 11060u_8u_6^3a_5^4a^2 + 660u_7^3a_5^3a^3 + 23790
  u_7^2u_6^2a_5^4a^2 \\
  &&+ 119040u_7u_6^4a_5^5a + 141400u_6^6a_5^6,
\end{eqnarray*}
\begin{eqnarray*}
\sigma^{(1)}&=&-u_{t,7}=-a^7 u_7-\frac{7}{2}a^6 a_5 u_6^2,\\
\gamma^{(1)}&=&\frac{\delta \rho^{(-1)}}{\delta u}=-D^5\frac{\partial a^{-1}}{\partial u_5},\\
\sigma^{(2)}&=&-u_{t,5}=-\frac{1}{2}\frac{a^2}{a_5}, \\
\gamma^{(2)}&=&\frac{\delta \rho^{(1)}}{\delta u}=
D^6 \frac{\partial \rho^{(1)}}{\partial u_6}-D^5   \frac{\partial \rho^{(1)}}{\partial u_5},\quad \rho^{(1)}=
\left(\frac{1}{a}a_5^2 u_6^2\right).
\end{eqnarray*}
\begin{eqnarray*}
u_{t,5}&=&\frac{1}{2}\frac{a^6}{a_5}, \\
u_{t,7}&=&a^7 u_7+\frac{7}{2}a^6 a_5 u_6^2,\\
u_{t,11}&=&a^{11}u_{11}
 + 33 		a^{10}	a_5 	u_{10}	u_6
 + 77		a^{10}	a_5		u_{9}	u_7
 + 99/2		a^{10}	a_5		u_8^2
 + 682		a^9		a_5^2	u_9		u_6^2\\
&& + 2574		a^9		a_5^2	u_8		u_7	u_6
 + 1892/3	a^9		a_5^2	u_7^3
 + 10098	a^8		a_5^3	u_8	u_6^3
 + 22066	a^8		a_5^3	u_7^2	u_6^2 \\
&& + 107525	a^7		a_5^4	u_7	u_6^4
 + 752675/6	a^6		a_5^5	u_6^6
\end{eqnarray*}

\subsection*
{Trivial $\rho^{(3)}$, base $b=4$}
\begin{eqnarray*}
R^{(5)}&=&	9	a^5 a_4	u_5,\\
R^{(4)}&=&	5	a^5	a_4	u_6			+ 34	a^4 a_4^2 u_5^2,\\
R^{(3)}&=&		a^5 a_4 u_7			+ 16    a^4 a_4^2 u_6 u_5   + 42 a^3 a_4^3 u_5^3 ,\\
R^{(2)}&=& - 4	a^4 a_4^2 u_7 u_5   - 56    a^3 a_4^3 u_6 u_5^2 - 140 a^2 a_4^4 u_5^4,\\
R^{(1)}&=&	2	a^4 a_4^2 u_8 		+ 2     a^4 a_4^2 u_7 u_6   + 52 a^3 a_4^3 u_7 u_5^2\\&&
          + 56  a^3 a_4^3 u_6^2 u_5 + 700   a^2 a_4^4 u_6 u_5^  + 1260 a a_4^5 u_5^5,\\
R^{(0)}&=&
- 2		a^4 	a_4^2 	u_9 u_5
- 2		a^4		a_4^2	u_7^2
- 56	a^3		a_4^3	u_8 u_5^2
- 156	a^3		a_4^3	u_7 u_6 u_5
- 1060	a^2		a_4^4	u_7 u_5^3 \\&&
- 1680	a^2		a_4^4	u_6^2 u_5^2
- 12180	a		a_4^5	u_6 u_5^4
- 17360			a_4^6	u_5^6
\end{eqnarray*}
\begin{eqnarray*}
	\sigma^{(1)}&=&-u_{t,7}=-\left( a^7 u_7+14 a^6 a_4 u_6 u_5 + 35 a^5 a_4^2 u_5^3\right),\\
	\gamma^{(1)}&=&\frac{\delta \rho^{(-1)}}{\delta u}= D^4\frac{\partial a^{-1}}{\partial u_4},\\
	\sigma^{(2)}&=&-u_{t,5}=-a^5 u_5, \\
	\gamma^{(2)}&=&\frac{\delta \rho^{(1)}}{\delta u}=
	-D^5 \frac{\partial \rho^{(1)}}{\partial u_5}+D^4   \frac{\partial \rho^{(1)}}{\partial u_4},\quad \rho^{(1)}=
	\left(\frac{1}{a}a_4^2 u_5^2\right).
\end{eqnarray*}
\begin{eqnarray*}
	u_{t,5}&=&a^5 u_5, \\
	u_{t,7}&=&a^7 u_7 + 14 a^6 a_4 u_6 u_5 + 35 a^5 a_4^2 u_5^3,\\
	u_{t,11}&=&a^{11}u_{11}
+ 44		a^{10}	a_4		u_{10}	u_5
+ 110		a^{10}	a_4		u_9		u_6
+ 176		a^{10}	a_4		u_8		u_7
+ 1144		a^9		a_4^2	u_9		u_5^2 \\&&
+ 5016		a^9		a_4^2	u_8		u_6 u_5
+ 3267		a^9		a_4^2	u_7^2	u_5
+ 4466		a^9		a_4^2	u_7		u_6^2
+ 21692		a^8		a_4^3	u_8		u_5^3 \\&&
+ 118184	a^8		a_4^3	u_7		u_6 u_5^2
+ 164560/3	a^8		a_4^3	u_6^3	u_5
+ 309485	a^7		a_4^4	u_7		u_5^4 \\&&
+ 871420	a^7		a_4^4	u_6^2	u_5^3
+ 3225750	a^6		a_4^5	u_6		u_5^5
+ 9784775/3	a^5		a_4^6	u_5^7
\end{eqnarray*}

\subsection*
{ Trivial $\rho^{(3)}$, base $b=3$}
\begin{eqnarray*}
R^{(5)}&=&	15 	a^5 a_3 u_4,\\
R^{(4)}&=&	14	a^5 a_3 u_5 + 115	a^4 a_3^2 u_4^2,\\
R^{(3)}&=&	6*	a^5 a_3 u_6	+ 129	a^4 a_3^2 u_5 u_4 + 450	a^3 a_3^3 u_4^3,\\
R^{(2)}&=&		a^5 a_3 u_7	+ 21	a^4 a_3^2 u_6 u_4 + 16	a^4 a_3^2 u_5^2
							+ 262   a^3 a_3^3 u_5 u_4^2 + 490 a^2 a_3^4 u_4^4,\\
R^{(1)}&=& - 2	a^4 a_3^2 u_7 u_4 - 2 a^4 a_3^2 u_6 u_5
           - 52 a^3 a_3^3 u_6 u_4^2 - 56 a^3 a_3^3 u_5^2 u_4
           - 700 a^2 a_3^4 u_5 u_4^3 - 1260 a a_3^5 u_4^5\\
R^{(0)}&=&	4	a^4 a_3^2 u_7 u_5  + 20 a^3 a_3^3 u_7 u_4^2 + 84 a^3 a_3^3 u_6 u_5 u_4
           + 420 a^2 a_3^4 u_6  u_4^3 + 56   a^3 a_3^3 u_5^3 \\&&
           + 1260 a^2 a_3^4 u_5^2 u_4^2 + 6720 a a_3^5 u_5 u_4^4 + 9100 a_3^6 u_4^6,
\end{eqnarray*}
\begin{eqnarray*}
	\sigma^{(1)}&=&-u_{t,7}=-\left(a^7 u_7 + 21 a^6 a_3 u_6 u_4
	+14   a^6  a_3  u_5^2
	+245  a^5  a_3^2 u_4^2 u_5
	+455  a^4 a_3^3 u_4^4\right),\\
	\gamma^{(1)}&=&\frac{\delta \rho^{(-1)}}{\delta u}= -D^3\frac{\partial a^{-1}}{\partial u_3},\\
	\sigma^{(2)}&=&-u_{t,5}=-()a^5 u_5  + 5  a^4 a_3  u_4^2), \\
	\gamma^{(2)}&=&\frac{\delta \rho^{(1)}}{\delta u}=
	D^4 \frac{\partial \rho^{(1)}}{\partial u_4}-D^3   \frac{\partial \rho^{(1)}}{\partial u_3},\quad \rho^{(1)}=
	\left(\frac{1}{a}a_3^2 u_4^2\right).
\end{eqnarray*}
\begin{eqnarray*}
	u_{t,5}&=&a^5 u_5  + 5  a^4 a_3  u_4^2  	, \\
	u_{t,7}&=& a^7 u_7 + 21 a^6 a_3 u_6 u_4
	+14   a^6  a_3  u_5^2
	+245  a^5  a_3^2 u_4^2 u_5
	+455  a^4 a_3^3 u_4^4,\\
	u_{t,11}&=&a^{11}u_{11}
	+55			a^{10}	a_3		u_{10}  u_4
	+154		a^{10}	a_3		u_9 u_5
	+286		a^{10}	a_3		u_8 u_6
	+176		a^{10}	a_3		u_7^2		\\&&
	+1760		a^{9}	a_3^2	u_9 u_4^2
	+8844		a^{9}	a_3^2	u_8 u_5 u_4
	+14014		a^{9}	a_3^2	u_7 u_6 u_4
	+9482		a^{9}	a_3^2	u_7 u_5^2	\\&&
	+12199		a^{9}	a_3^2	u_6^2 u_5	
	+41140		a^{8}	a_3^3	u_8 u_4^3	
	+268532		a^{8}	a_3^3	u_7 u_5 u_4^2
	+173723		a^{8}	a_3^3	u_6^2 u_4^2	\\&&
	+476850		a^{8}	a_3^3	u_6 u_5^2 u_4
	+164560/3	a^{8}	a_3^3	u_5^4
	+743325		a^{7}	a_3^4	u_7 u_4^4
	+5344460	a^{7}	a_3^4	u_6 u_5 u_4^3	\\&&
	+11133980/3	a^{7}	a_3^4	u_5^3 u_4^2
	+10343905	a^{6}	a_3^5	u_6 u_4^5
	+36171410	a^{6}	a_3^5	u_5^2 u_4^4	\\&&
	+320101925/3a^{5}	a_3^6	u_5 u_4^6
	+283758475/3a^{4}	a_3^7	u_48
\end{eqnarray*}

\section*{6. Results and Discussion}

 We obtained the ``top level" parts of seemingly new hierarchies up to their dependencies on the ``top order" derivative that is present in the separant.
  The dependencies on $u$, $u_1$, $\dots$, $u_{b-1}$ could'nt be solved completely despite numerous attempts for attacking this problem suggesting that one should use transformations to eliminate some arbitrary functions.

  Assuming no lower order dependencies,  it is possible
   to show that the equations that we obtain can be mapped to polynomial equations.
  This is achieved by a sequence of potentiations, generalized contact transformations and point transformations.
  We recall that a potentiation is a special type of Miura map, defined by $v=u_1$. Then, if $u_t=F$, then $v_t=DF$.  If $F$ has no dependency on $u$, $DF$ will be a local function of $v$ and its derivatives. If $F$ has a non-polynomial dependency on $u_j$, $j\le b$, then $v_t$ will be polynomial in $v_b=u_{b+1}$, hence, in the level grading terminology, the base level will decrease.  This potentiation procedure can be continued until the base level is $b=0$, that means the evolution equation is polynomial in $u_j$, $j>0$ but has non-polynomial dependency on $u$ only. This procedure has been applied to non-quasilinear equations of orders $3$ and $5$ and corresponding equations in our list have been obtained.

  The next step in the reduction procedure is to apply the generalized contact transformation given by \cite{MSS1991},
  $$dx'=\rho dx + \sigma dt,\quad u'=u, \quad t'=t, \quad \rho_t=D\sigma, \quad \rho{\not\in} Im(D),$$
 that implies
 $$u'_{t'}= u_t-(\sigma/\rho)\ u_1,  \quad\quad u'_ {k'}=(\rho^{-1}D)^k \ u.$$
If there is a non-trivial conserved density depending on $u$ and $u_1$, this is generalized contact transformation maps $u_t$ to an equation where the separant is equal to $1$.  In our case, since we assumed that
 the canonical density $\rho^{(-1)}$ is non-trivial, we applied this transformation with $\rho^{(-1)}$ depending on $u$ only, to set the separant equal to $1$ but at this stage we still had non-polynomial dependencies.

  The final step in the sequence of transformations is a point transformation  $\tilde{u}=\phi(u)$, aiming to eliminate
  the dependency on $u_{m-1}$.  Since our evolution equations are now of the form
  $$u_t=u_m +B(u) u_1 u_{m-1}+\dots,$$
  it can be seen that
  $$\tilde{u}_t -\tilde{u}_m=\phi_u(u_m+B(u) u_1 u_{m-1}+\dots)  -
                        (\phi_uu_m+m\phi_{u,u} u_1 u_{m-1}+\dots).
   $$
  Thus, $u_{m-1}$ dependency is eliminated by choosing
  $$m\phi_{u,u}=B(u)\phi_u.$$
  The existence of local solutions to this equation depends on the triviality of the canonical density $\rho^{(0)}=\frac{\partial{F}/\partial{u_{m-1}}}{\partial{F}/\partial{u_m}}$, that holds in our case.  This transformation eliminates $u_{m-1}$ and in our case, it reduces all equation to polynomial equations that are in fact independent of $u$ also.

 Based on these top level-top order classification and transformation results, we conjecture that all scalar evolution equations in $1$ space dimension, integrable in the sense of admitting a formal symmetry, are symmetries of a polynomial equation of order $3$ or $5$.

\section*{Appendix A: Partitions of integers}

The partitions of integers are computed using the Matlab function  ``partitions.m" written by
 John D'Errico \cite{Errico}.
The command \texttt{  partitions(n)} produces partitions of the integer $n$ as a matrix $P_n$.  Each row of the matrix $P_n$ corresponds to a different partition of $n$,
A value $k$  at the entry $(i,j)$ means that, in the $i$th partition,  the integer $j$ occurs $k$ times.  For example the rows of $P_4$ correspond respectively to the following  partitions of  $4$.
$$1+1+1+1,\quad 1+1+2,\quad 2+2,\quad 1+3,\quad 4.$$
We present below partitions of integers $n\le 7$.

$$
P_1=\left[ {\begin{array}{c} 1\end{array}}\right],\quad
P_2=\left[ {\begin{array}{cc}
		2 & 0 \\
		0 & 1      \end{array} } \right],\quad
P_3=\left[ {\begin{array}{ccc}
	3 & 0 & 0 \\
	1 & 1 & 0 \\
	0 & 0 & 1     \end{array} } \right],\quad
P_4=\left[ {\begin{array}{cccc}
4 &    0&     0&     0\\
2 &    1&     0&     0\\
0 &    2&     0&     0\\
1 &    0&     1&     0\\
0 &    0&     0&     1	 \end{array} } \right],\quad
$$

$$
P_5=\left[ {\begin{array}{ccccc}
5&     0&     0&     0&     0\\
3&     1&     0&     0&     0\\
1&     2&     0&     0&     0\\
2&     0&     1&     0&     0\\
0&     1&     1&     0&     0\\
1&     0&     0&     1&     0\\
0&     0&     0&     0&     1 	\end{array}}\right],\quad
P_6=\left[ {\begin{array}{cccccc}
6&     0&     0&     0&     0&     0\\
4&     1&     0&     0&     0&     0\\
2&     2&     0&     0&     0&     0\\
0&     3&     0&     0&     0&     0\\
3&     0&     1&     0&     0&     0\\
1&     1&     1&     0&     0&     0\\
0&     0&     2&     0&     0&     0\\
2&     0&     0&     1&     0&     0\\
0&     1&     0&     1&     0&     0\\
1&     0&     0&     0&     1&     0\\
0&     0&     0&     0&     0&     1\end{array}}\right],\quad	
P_7=\left[ {\begin{array}{ccccccc}
7&     0&     0&     0&     0&     0&     0\\
5&     1&     0&     0&     0&     0&     0\\
3&     2&     0&     0&     0&     0&     0\\
1&     3&     0&     0&     0&     0&     0\\
4&     0&     1&     0&     0&     0&     0\\
2&     1&     1&     0&     0&     0&     0\\
0&     2&     1&     0&     0&     0&     0\\
1&     0&     2&     0&     0&     0&     0\\
3&     0&     0&     1&     0&     0&     0\\
1&     1&     0&     1&     0&     0&     0\\
0&     0&     1&     1&     0&     0&     0\\
2&     0&     0&     0&     1&     0&     0\\
0&     1&     0&     0&     1&     0&     0\\
1&     0&     0&     0&     0&     1&     0\\
0&     0&     0&     0&     0&     0&     1\end{array}}\right].	
$$

\section*{Appendix B: Level homogeneous polynomials}
A partition of the integer $n$ given by the $i$th row of the matrix $P_n$ corresponds  to a monomial of level $n$ above a base level $b$ as follows.  Each non-zero value $k_{i,j}$ at the $(i,j)$ entry of the matrix corresponds to a term $u_{b+j}^{k_{i,j}}$. Since the sum of $i$ times the values in the $i$th row is $n$ the product of the corresponding terms is a monomial of level $n$ above the base level $b$. As an example,
we have the following correspondences for $n=4$.
\begin{eqnarray*}
1+1+1+1&:&u_{b+1}^4,\\
1+1+2  &:&u_{b+1}^2u_{b+2},\\
2+2    &:&u_{b+2}^2,\\
1+3    &:&u_{b+1}u_{b+3},\\
4      &:&u_{b+4}.
\end{eqnarray*}
We transfer the matrices $P_n$  above to level homogeneous polynomials in  REDUCE format as follows.

\begin{verbatim}
t:=P4;ncol:=4;nrow:=5;					% t=matrix of partions in REDUCE format
ubb:=mat((ubp1,ubp2,ubp3,ubp4));		% derivatives of order b+1,b+2,b+3 b+4
kat:=tp(mat((k01,k02,k03,k04,k05)));    % coefficinets of the polynomial
term:=0;for i:=1 step 1 until nrow do
<<terma:=1;for j:=1 step 1 until ncol do <<terma:=terma*ubb(1,j)**t(i,j)>>;
term:=term+kat(i,1)*terma>>;			% result is below
utbmm04:=
k01*ubp1**4 + k02*ubp1**2*ubp2 + k03*ubp2**2 + k04*ubp1*ubp3 + k05*ubp4;
\end{verbatim}
The general form of the top level parts are given below.
\begin{eqnarray*}
	m-b=1&:&	\{ u_{b+1}    \}\\
	m-b=2&:&\{u_{b+1}^2,u_{b+2}    \}\\
	m-b=3&:&\{u_{b+1}^3,u_{b+1}u_{b+2},u_{b+3}    \}\\
	m-b=4&:&\{u_{b+1}^4,u_{b+1}^2u_{b+2},u_{b+2}^2,u_{b+1}u_{b+3},u_{b+4}    \}\\
	m-b=5&:&\{u_{b+1}^5,u_{b+1}^3u_{b+2},u_{b+1}u_{b+2}^2,u_{b+1}^2u_{b+3},u_{b+2}u_{b+3},u_{b+1}u_{b+4},u_{b+5}    \}\\
	m-b=6&:&\{u_{b+1}^6,u_{b+1}^4u_{b+2},u_{b+1}^2u_{b+2}^2,u_{b+2}^3,u_{b+1}^3u_{b+3},u_{b+1}u_{b+2}u_{b+3},u_{b+3}^2,u_{b+1}^2u_{b+4},\\
	&&  u_{b+2}u_{b+4},u_{b+1}u_{b+5},u_{b+6}  \}\\
	m-b=7&:&\{u_{b+1}^7,u_{b+1}^5u_{b+2},u_{b+1}^3u_{b+2}^2,u_{b+1}u_{b+2}^3,u_{b+1}^4u_{b+3},u_{b+1}^2u_{b+2}u_{b+3},u_{b+2}^2u_{b+3},u_{b+1}u_{b+3}^2,\\
	&& u_{b+1}^3u_{b+4},u_{b+1}u_{b+2}u_{b+4},u_{b+3}u_{b+4},u_{b+1}^2u_{b+5},u_{b+2}u_{b+5},u_{b+1}u_{b+6},u_{b+7}  \}\\
	m-b=8&:&\{u_{b+1}^8,u_{b+1}^6u_{b+2},u_{b+1}^4u_{b+2}^2,u_{b+1}^2u_{b+2}^3,u_{b+2}^4,u_{b+1}^5u_{b+3},u_{b+1}^3u_{b+2}u_{b+3}, u_{b+1}u_{b+2}^2u_{b+3},\\
	&& u_{b+2}^2u_{b+3}^2,u_{b+2}u_{b+3}^2,u_{b+1}^4u_{b+4},u_{b+1}^2u_{b+2}u_{b+4},u_{b+2}^2u_{b+4},u_{b+1}u_{b+3}u_{b+4},u_{b+4}^2,\\
	&& u_{b+1}^3u_{b+5},u_{b+1}u_{b+2}u_{b+5},u_{b+3}u_{b+5},u_{b+1}^2u_{b+6},u_{b+2}u_{b+6},u_{b+1}u_{b+7},u_{b+8}  \}\\
	m-b=9&:&\{u_{b+1}^9,u_{b+1}^7u_{b+2},u_{b+1}^5u_{b+2}^2,u_{b+1}^3u_{b+2}^3,u_{b+1}u_{b+2}^4,u_{b+1}^6u_{b+3},u_{b+1}^4u_{b+2}u_{b+3},\\
	&& u_{b+1}^2u_{b+2}^2u_{b+3},u_{b+2}^3u_{b+3},u_{b+1}^3u_{b+3}^2,u_{b+1}u_{b+2}u_{b+3}^2,u_{b+3}^3,u_{b+1}^5u_{b+4},u_{b+1}^3u_{b+2}u_{b+4},\\
	&& u_{b+1}u_{b+2}^2u_{b+4},u_{b+1}^2u_{b+3}u_{b+4},u_{b+2}u_{b+3}u_{b+4},u_{b+1}u_{b+4}^2,u_{b+1}^4u_{b+5},u_{b+1}^2u_{b+2}u_{b+5},\\
	&& u_{b+2}^2u_{b+5},u_{b+1}u_{b+3}u_{b+5},u_{b+4}u_{b+5},u_{b+1}^3u_{b+6},u_{b+1}u_{b+2}u_{b+6},u_{b+3}u_{b+6},u_{b+1}^2u_{b+7},\\
	&& u_{b+2}u_{b+7},u_{b+1}u_{b+8},u_{b+9}  \}\\
	m-b=10 &:&\{u_{b+1}^{10},u_{b+1}^8u_{b+2},u_{b+1}^6u_{b+2}^2,u_{b+1}^4u_{b+2}^3,u_{b+1}^2u_{b+2}^4,u_{b+2}^5,u_{b+1}^7u_{b+3},\\
	&& u_{b+1}^5u_{b+2}u_{b+3},u_{b+1}^3u_{b+2}^2u_{b+3},u_{b+1}u_{b+2}^3u_{b+3},u_{b+1}^4u_{b+3}^2,u_{b+1}^2u_{b+2}u_{b+3}^2,\\
	&& u_{b+2}^2u_{b+3}^2,u_{b+1}u_{b+3}^3,u_{b+1}^6u_{b+4},u_{b+2}^3u_{b+4},u_{b+1}^4u_{b+2}u_{b+4},u_{b+1}^2u_{b+2}^2u_{b+4},\\
	&& u_{b+1}^3u_{b+3}u_{b+4},u_{b+1}u_{b+2}u_{b+3}u_{b+4},u_{b+3}^2u_{b+4},u_{b+1}^2u_{b+4}^2,u_{b+2}u_{b+4}^2,u_{b+1}^5u_{b+5}, \\
	&& u_{b+1}^3u_{b+2}u_{b+5},u_{b+1}u_{b+2}^2u_{b+5},u_{b+1}^2u_{b+3}u_{b+5},u_{b+2}u_{b+3}u_{b+5},u_{b+1}u_{b+4}u_{b+5},u_{b+5}^2,\\
	&& u_{b+1}^4u_{b+6},u_{b+1}^2u_{b+2}u_{b+6},u_{b+2}^2u_{b+6},u_{b+1}u_{b+3}u_{b+6},u_{b+4}u_{b+6},u_{b+1}^3u_{b+7},\\
	&& u_{b+1}u_{b+2}u_{b+7},u_{b+3}u_{b+7},u_{b+1}^2u_{b+8},u_{b+2}u_{b+8},u_{b+1}u_{b+9},u_{b+10}\}\end{eqnarray*}

\section*{Appendix C: Sample REDUCE programs}

In our REDUCE programs, we define the $i$th partial derivative of $u$ with respect to $x$ as an indeterminate \texttt{ui}. Then, for functions depending on the derivatives of $u$,  we define  the total derivative of functions depending on at most say $30$th derivative of $u$ by the REDUCE procedure
\begin{verbatim}
procedure tdf(var);tdf(var)=df(var,u30)*u31+...+df(var,u)*u1}.
\end{verbatim}
Integration by parts is done by the procedure \texttt{ intk},
\begin{verbatim}
procedure intk(var,um,umm1,k)$
var-tdf(coeffn(coeffn(var,um,1),umm1,k)*umm1^(k+1)/(k+1))$
\end{verbatim}
where, $um$ and $umm1$ denote respectively $u_m$ and $u_{m-1}$,  the monomial   {\it $u_m u_{mm1}^k$} is the  top order derivative of the polynomial \texttt{var}.  We apply this procedure repeatedly until the top order monomial is nonlinear in its highest order.

REDUCE program for the computation of the recursion operator	
\begin{verbatim}
%%%%%%%%%%%%%%%%%%%%%%%%%%%%%%%%%%%%%%%%%%%%%%%%%%%%%%%%%%%%%%%%%%%%%
% find the recursion operator
% find the integrals
eqn:=gamma*ut3$itg:=0$eqn$
coeffn(ws,u6,1)$coeffn(ws,u5,0)$itg:=itg+ws*u5**1/1$eqn-tdf(itg)$
coeffn(ws,u5,1)$coeffn(ws,u4,1)$itg:=itg+ws*u4**2/2$eqn-tdf(itg)$
coeffn(ws,u5,1)$coeffn(ws,u4,0)$itg:=itg+ws*u4**1/1$eqn-tdf(itg)$
itg3:=itg$
% The recursion operator of order 2
depend {mm1,mm2,mm3,mm4},u3$
%%%%%%%%%%%%%%%%%%%%%%%%%%%%%%%%%%%%%%%%%%%%%%%%%%%%%%%%%%%%%%%%%%
% R acting on each flow are given below
% R should have level 2;
% gamma = euler derivative of rhom1 has level 3
% R=a^2 D^2 + (mm1*u4) *D +(mm2*u5+mm3*u4**2)+ mm4* D^(-1)* gamma
%%%%%%%%%%%%%%%%%%%%%%%%%%%%%%%%%%%%%%%%%%%%%%%%%%%%%%%%%%%%%%%%%%
rrr3:=a**2*tdf(tdf(ut3))+(mm1*u4)*tdf(ut3)+(mm2*u5+mm3*u4**2)*ut3+mm4*itg3$
rrr5:=a**2*tdf(tdf(ut5))+(mm1*u4)*tdf(ut5)+(mm2*u5+mm3*u4**2)*ut5+mm4*itg5$
rrr7:=a**2*tdf(tdf(ut7))+(mm1*u4)*tdf(ut7)+(mm2*u5+mm3*u4**2)*ut7+mm4*itg7$
%%%%%%%%%%%%%%%%%%%%%%%%%%%%%%%%%%%%%%%%%%%%%%%%%%%%%%%%%%%%%%%%%%%
% R acting on each flow should give the next one
% We allow for a constant
temp3:=rrr3-lan5*ut5;
temp5:=rrr5-lan7*ut7;
%%%%%%%%%%%%%%%%%%%%%%%%%%%%%%%%%%%%%%%%%%%%%%%%%%%%%%%%%%%%%%%%%%%%%%%%%%%%%%%%%%
\end{verbatim}
\section*{Appendix D: Sample REDUCE program the commutativity of the flows}
We show that: (a) the equations $u_{t,5}, u_{t,7}, u_{t,9}, u_{t,11}, u_{t,13}, u_{t,15}$ form a commuting flow with $u_{t,3}$.
 (b)the equations $u_{t,7}, u_{t,9}, u_{t,11}, u_{t,13}, u_{t,15}$ form a commuting flow with $u_{t,5}$.
 (c)the equations $u_{t,9}, u_{t,11}, u_{t,13}$ and $u_{t,15}$ form a commuting flow with $u_{t,7}$.
 (d)the equations $u_{t,11}, u_{t,13}$ and $u_{t,15}$ form a commuting flow with $u_{t,9}$.
 (e)the equations $u_{t,13}, u_{t,15}$ form a commuting flow with $u_{t,11}$.
 (f)the equations $u_{t,15}$ form a commuting flow with $u_{t,13}$.
 Below we give the program for part (a) for equations  $u_{t,5}, u_{t,7}, u_{t,9}$ only.
\begin{verbatim}
%%%%%%%%%%%%%%%%%%%%%%%%%%%%%%%%%%%%%%%%%%%%%%%%%%%%%%%%%%%%%%%
% Show that the flow is commuting
% Show that the equations ut5, ut7,ut9, are symmetries of ut3
% We need k times the total derivative of ut3
ff:=ut3$
ffx1:=tdf(ff)$
ffx2:=tdf(ffx1)$
...
ffx15:=tdf(ffx14)
%%%%%%%%%%%%%%%%%%%%%%%%%%%%%%%%%%%%%%%%%%%%%%%%%%%%%%%%%%%%%%%
sigma:=ut5$
denk:=df(sigma,u5)*ffx5+df(sigma,u4)*ffx4+df(sigma,u3)*ffx3
      -df(ff,u3)*tdf(tdf(tdf(sigma)));pause;
sigma:=ut7$
denk:=df(sigma,u7)*ffx7+df(sigma,u6)*ffx6
      +df(sigma,u5)*ffx5+df(sigma,u4)*ffx4+df(sigma,u3)*ffx3
      -df(ff,u3)*tdf(tdf(tdf(sigma)));pause;
sigma:=ut9$
denk:=df(sigma,u9)*ffx9+df(sigma,u8)*ffx8
      +df(sigma,u7)*ffx7+df(sigma,u6)*ffx6
      +df(sigma,u5)*ffx5+df(sigma,u4)*ffx4+df(sigma,u3)*ffx3
      -df(ff,u3)*tdf(tdf(tdf(sigma)));pause;
%%%%%%%%%%%%%%%%%%%%%%%%%%%%%%%%%%%%%%%%%%%%%%%%%%%%%%%%%%%%%%%%%%%%%
\end{verbatim}

\end{document}